\documentclass{PoS}
\usepackage{tensor}
\usepackage{mathtools}
\usepackage{amstext,amsmath,amssymb,amsfonts,bbm}
\usepackage[latin1]{inputenc}
\usepackage{fancyhdr}

\usepackage{verbatim}
\usepackage{epigraph}
\usepackage{url}
\usepackage{footmisc}
\usepackage{physics}

\usepackage{todonotes}

\usepackage{wasysym} 
\usepackage{marvosym} 
\usepackage{textcomp} 

\newcommand{\beq}{\begin{equation}}
\newcommand{\eeq}{\end{equation}}
\newcommand{\be}{\begin{equation}}
\newcommand{\bee}{\begin{equation}}
\newcommand{\ee}{\end{equation}}
\newcommand{\bea}{\begin{eqnarray}}
\newcommand{\eea}{\end{eqnarray}}

\newcommand{\sign}{{\rm sign}}

\newcommand{\bal}{\begin{align}}
\newcommand{\eal}{\end{align}}

\newcommand{\cJ}{{\cal{J}}}

\title{The Tensor Track V: Holographic Tensors}

\ShortTitle{Holographic Tensors}

\author{Nicolas Delporte and \speaker{Vincent Rivasseau}\\
        Laboratoire de physique th\'eorique,
        CNRS, Univ. Paris-Sud, Universit\'e Paris-Saclay, 91405 Orsay, France\\
        E-mails: \email{delporte.nico@gmail.com, vincent.rivasseau@u-psud.fr}}

\abstract{We review the fast developing subject of tensor models for the NAdS$_2$/NCFT$_1$ holographic correspondence. 
We include a brief review of the Sachdev-Ye-Kitaev (SYK) model and then focus on the associated quantum mechanical tensor models (GW and CTKT). We examine their main features and how they compare with SYK. 
To end, we discuss different extensions: the large $D$ limit of matrix-tensor models, the large $N$ expansion of symmetric/antisymmetric tensors, the use of probes, the construction of a bilocal action for tensors, some attempts to extend the above models to higher dimensions and a proposal to break the tensor symmetry.}

\FullConference{Corfu Summer Institute 2017 'School and Workshops on Elementary Particle Physics and Gravity'\\
		2-28 September 2017\\
		Corfu, Greece}

\begin{document}

\section{Introduction}

The tensor track \cite{tensortrack} is an ongoing program which proposes to use random tensors \cite{tensors}
to progress on quantizing gravity. In this new issue of the franchise we focus on a brief and
necessarily incomplete review of the most remarkable development of the last two years, 
namely the discovery and burgeoning study of \emph{holographic tensor models}. 
%


We would like quantum gravity to provide some light on early big-bang cosmology and on the quantum
information paradox for black holes. Holography (and in particular the AdS/CFT correspondence \cite{ADSCFT}) provides 
an effective definition of quantum gravity systems dual to certain conformal field theories. It raised many hopes 
that quantum field theory, in a broad sense, may be used to understand quantum gravity. However until recently 
the lack of simple solvable examples of this correspondence prevented to extract easily the gravitational content. 

String theory provides a microscopic theory of supersymmetric extremal black holes. 
Nevertheless, as often remarked, such black-holes do not radiate since their temperature is exactly zero.
As a consequence many experts disagree on how to solve the so-called quantum information paradox. Are there firewalls or not at the horizon crossing?

More generally there has been in recent years a growing desire to forge mathematical tools 
that can go beyond the current limitations of superstring theory and of standard
AdS/CFT correspondence. We would like such tools not to depend completely
on supersymmetry. Indeed whether supersymmetry is a fundamental law of nature or not, it has to be certainly
broken or absent in many physical situations, in particular for astrophysical black holes. We would like
to quantize branes \emph{ab initio}, not just as subsectors constrained by string theory. 
We would like also to give a meaning to some kind of functional 
integral over space-times, pondered by an action of the Einstein-Hilbert type. 
This may allow to understand quantum gravity in any dimension and for non-AdS spaces. A byproduct could be
a hopefully simple "mean field" theory of quantum gravity in large dimensions.

This wish list may seem made of distant dreams. However random tensors models
are an interesting tool to start filling the bill. Their definition does not require supersymmetry,
they are not limited to a particular rank/dimension, and they provide a sum over geometries not limited
to AdS.

The world of random tensors has a surprisingly simple entrance door, namely the
family of melonic graphs \cite{Bonzom:2011zz}. Beyond this modest  door
lies a marvelous world of dazzling complexity. For instance the $1/N$ tensor expansion \cite{1/N} of even the simplest
random tensor models provides a hierarchy for  the sum over the huge category of piecewise linear manifolds. This
is true in any rank/dimension. In particular in dimension/rank four, tensor models 
can distinguish all different smooth structures over all triangularizable manifolds, which
remain today poorly understood. Indeed there is yet no full classification
of such structures  even for a \emph{single} compact four dimensional topological manifold.
For the sphere the issue corresponds to the so-called smooth Poincar\'e conjecture. There is also no good enumeration of
general triangulations of the sphere in dimension greater than two. Any new mathematical tool in this area is welcome. It may
bring also new questions, possibly both more relevant for quantum gravity and easier to answer. For instance we argued that since gravity is not topological in higher dimensions, what should be relevant for quantum gravity is the enumeration of triangulations of given \emph{Gurau degree} rather than of fixed topology.

The survey is organised as follows. In section \ref{sec : syk}, we will recall the computation of the two and four point function of SYK and what it aimed at, mainly following \cite{MS2016}. In section \ref{sec : holographic tm}, after a brief comparison of the well established $1/N$ limits, we present the two leading holographic tensor models and (a portion of) what has been achieved up to now (spectra, diagrams of subleading orders, etc.). We end in section \ref{sec : further issues} with a selection of different developments, exploring the vast landscape of tensor models and looking for further intersections with the holographic duality.

\section{Blitz Review of the SYK Model}
\label{sec : syk}

In March 2015 Maldacena, Shenker and Stanford \cite{Maldacena:2015waa} established
that for a many-body quantum system at temperature $T$ the Lyapunov exponent for transient chaos 
in a four point correlator maximally spaced on the thermal circle
\bee
F(t) = \Tr [yV yW (t)yV yW (t)], \quad y := Z^{-1/4} e^{- \beta H /4},
\ee
in any quantum system is bounded by $\lambda_L \le 2 \pi T / \hbar$ (hereafter the MSS bound). This bound
is valid under very general assumptions (analyticity in a strip of width $\beta/2$ in the complex time plane
and reasonable behavior at infinity). 
More precisely they found that, noting $N$ the number of degrees of freedom in the system
and $t_d$ and $t_s$ its diffusion and scrambling times\footnote{The first characterizing the response to a local perturbation in a many-body system (and obtained with time-ordered correlators), whereas the second measures the spreading of quantum information across the whole system.}, one should expect
\bee F(t) \simeq  \left(a -\frac{b}{N^2} e^{\lambda_L t}\right)^{-b}, 
\quad t_d <  t < t_s ,
\ee
where the maximal Lyapunov exponent $\lambda_L$ must obey $\lambda_L  \le 2 \pi T /  \hbar$.
Furthermore they argued convincingly that saturation of this bound is a strong indication of the presence of a gravitational dual. Indeed saturation implies some Regge trajectory for a spin two particle or "graviton". 

Almost simultaneously in 2015 Kitaev found a very simple model which saturates the MSS bound, indicating  the surprising presence of a gravitational dual in two dimensions \cite{SYK}. It is a quasi-conformal one dimensional quantum mechanics model with action
\bee I = \int dt  \biggl( \frac{i}{2} \sum_{i} \psi_i \frac{d}{dt}\psi_i  -i^{q/2}	 \sum_{1 \le i_1 <  \cdots < i_q \le N}	J_{i_1, \cdots , i_q}	\psi_{i_1} \cdots \psi_{i_q}\biggr)
\ee
with $J_I$ a quenched iid random tensor ($<J_I J_{I'}> = \delta_{II'}J^2 (q-1)! N^{-(q-1)}$), and $\psi_i$ an $N$-vector Majorana Fermion.

This model now called the Sachdev-Ye-Kitaev model or \emph{SYK} 
is solvable as $N \to \infty$, being approximately reparametrization invariant (i.e. conformal)
in the infrared limit. Moreover an attentive study of the four point function reveals that the model
saturates the MSS bound \cite{SYK,MS2016}.
The corresponding so-called NAdS$_2$/NCFT$_1$ (where N stands for ``near") holography 
attracted considerable interest and is currently the subject of active investigation.

The model came somewhat as a surprise since solvability and chaotic behavior were previously somehow considered as incompatible.
It raises indeed many new questions, especially  since quantum gravity in two dimensions ought to be \emph{topological}. 
Nevertheless any near-extremal blackhole should have a two dimensional "throat" in which the radial distance to the horizon and the time should be the effective interesting dimensions.
Therefore simple models of the NAdS$_2$/NCFT$_1$ correspondence are particularly interesting, as they could 
shed light on issues such as the information paradox.

As mentioned above, the reason the theory can be \emph{solved}
in the limit $N \to \infty$ is because the leading Feynman graphs of the SYK model are the \emph{melonic} graphs \cite{Bonzom:2011zz}
which dominate the $1/N$ tensor expansion \cite{1/N}. We give now 
a brief review of the corresponding computation of the 2 and 4 point function in the infrared limit.

\subsection{Two Point Function}
\begin{figure}
\centerline{\includegraphics[width=12cm]{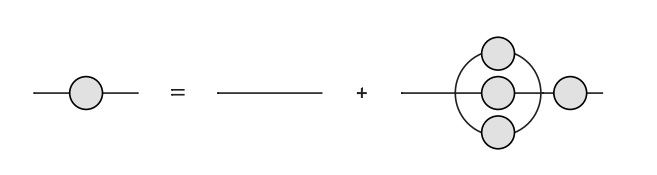}}
\caption{Two-Point Melonic Function}
\label{fig : 2 point melonic}
\end{figure}
In order to build a dimensionless parameter to describe different regimes of the theory, one can use $\beta J$, the UV-dimension of $J$ being $1$. Then a low-temperature (or infrared) regime, at fixed coupling, can be as well described by a strong coupling regime at fixed temperature, or $\beta J >> 1$ (finite temperature or not will be distinguished by the domain of 
the Euclidean time $\tau$). 

To start, we need to obtain the Euclidean 2-point function 
\bee
G(\tau) = \frac{1}{N}\sum_i\expval{\psi_i(\tau)\psi_i(0)}
\ee
or (by abuse of notation) its Fourier transform
\bee
G(\omega) = \int_{-\infty}^\infty \dd{\tau} e^{i\omega \tau}G(\tau).
\ee
At finite temperature, the (Matsubara) frequencies are quantized: $\omega_n= \frac{2\pi }{\beta} (n + 1/2)$ and the Euclidean time is bounded $0\leq \tau\leq \beta$. It is also convenient to define the quantities $\Delta = 1/q$, $\cJ^2 = \frac{qJ^2}{2^{q-1}}$.

At leading order in $1/N$, with the free propagator $G_0^{-1} (\omega) = i \omega$ and the self-energy $\Sigma(\omega)$, the Schwinger-Dyson equations in the large $N$ (melonic) limit read
\bee G^{-1} (\omega) = G_0^{-1} (\omega) - \Sigma (\omega),\quad  \Sigma (\tau) = J^2 G(\tau )^{q-1} ,
\ee
that is depicted in Fig. \ref{fig : 2 point melonic}. The ``blob" indicates a full propagator and a quenched average was taken.
The first equation is the usual one linking the complete 2-point function to the self-energy.
Then, taking advantage of the form of the free propagator, the IR limit simplifies the above equation to the simpler
\bee 
\label{eq:convolution}
\int \dd \tau^\prime J^2 G(\tau-\tau^\prime) G(\tau^\prime - \tau^{\prime \prime} )^{q-1} = -\delta (\tau - \tau^{\prime\prime}).
\ee
Reparametrization invariance of eq. \refeq{eq:convolution} under any differentiable function $f$:
\bee G (\tau, \tau') \rightarrow [ f'(\tau ) f'(\tau ') ]^\Delta  G(f(\tau), f(\tau ')), \quad  \Sigma (\tau, \tau') \rightarrow [ f'(\tau ) f'(\tau ') ]^{\Delta (q-1)}  \Sigma(f(\tau), f(\tau ')),
\ee
suggests to search for a particular solution of type
\bee \label{2point}
G_c (\tau)  = b \vert \tau \vert^{-2 \Delta}  {\rm sign} \ \tau, \quad
J^2 b^q \pi = \left(\frac{1}{2} - \Delta\right) \tan (\pi \Delta ) .
\ee 
 The equation for $b$ comes from the formula
\bee \int_{- \infty}^{+ \infty} d \tau e^{i \omega \tau}  {\rm sign}  \tau \vert \tau \vert^{-2 \Delta} = 2^{1 - 2 \Delta} i \sqrt{\pi} \frac{\Gamma (1 - \Delta )}{\Gamma (\frac{1}{2} + \Delta)} \vert \omega \vert^{2 \Delta -1}  {\rm sign} \ \omega .
\ee
Applying reparametrization $f_\beta ( \tau) = \tan (\pi\tau/\beta )$ leads to 
the finite temperature two-point function
\bee G_c ( \tau) = \left[\frac{\pi }{\beta \sin (\pi \tau / \beta)} \right]^{2 \Delta} b \ {\rm sign } \;\tau.
\ee
Recalling that $\Delta = 1/q$,  the anomalous dimension for $ G_c (\tau)  \propto  \tau^{-2/q}$ at small $\omega$
corresponds to the theory being just renormalizable in the tensor field theory sense \cite{tensortrack}.

It is important to notice that the reparametrization symmetry has been spontaneously broken by the solution \eqref{2point} to an $SL(2,\mathbb{R})$: 
$\tau \rightarrow f(\tau) = \frac{a \tau + b}{c\tau + d}$.

\subsection{Four Point Function}
The equation for the four-point function in the large $N$ limit is
 \begin{equation}
 \frac{1}{N^2}\sum_{1\leq i,j\leq N}\expval{T\left(\psi_i(\tau_1)\psi_i(\tau_2)\psi_j( \tau_3)\psi_j(\tau_4)\right)} =  G(\tau_{12})G(\tau_{34}) + \frac{1}{N} 
 \mathcal{F}(\tau_1,\tau_2, \tau_3, \tau_4) + \cdots
\end{equation}
where we write $\tau_{12} = \tau_1 - \tau_2$. 

\begin{figure}
\centerline{\includegraphics[width=12cm]{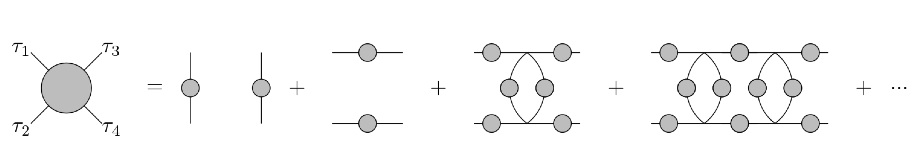}}
\caption{Four Point Melonic Function}
\end{figure}

The function $ \mathcal{F}$
then develops as a geometric series in \emph{rungs}. Calling $\mathcal{F}_n$ the ladder with $n$ ``rungs" we have
$ \mathcal{F} = \sum_{n \ge 0}  \mathcal{F}_n$, with $\mathcal{F}_0 = -G(\tau_{13})G(\tau_{24}) + G(\tau_{14})G(\tau_{23})$. The induction rule is
\bea \nonumber
\mathcal{F}_{n+1} (\tau_1,\tau_2, \tau_3, \tau_4) &=&  
\int d \tau d \tau^\prime K(\tau_1,\tau_2, \tau, \tau')  \mathcal{F}_{n} (\tau,\tau', \tau_3, \tau_4) .
\eea
The rung operator $K$ adds one rung to the ladder. It acts on the space of ``bilocal" functions
with kernel
\bee K(\tau_1,\tau_2, \tau_3, \tau_4) = -J^2(q-1)G(\tau_{13})G(\tau_{24})G(\tau_{34})^{(q-2)}.
\ee
\begin{figure}
\centerline{\includegraphics[width=4cm]{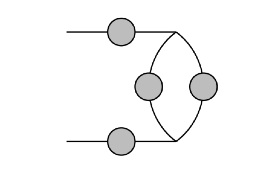}}
\caption{The Rung Operator}
\end{figure}

The geometric series gives:
\bee   \mathcal{F} = \sum_{n \ge 0}  \mathcal{F}_n = 
\sum_{n \ge 0} K^n \mathcal{F}_0 = \frac{1}{1-K} \mathcal{F}_0 
\ee
or, more explicitly 
\bee
\mathcal{F}(\tau_1,\tau_2, \tau_3, \tau_4) = \int d \tau d \tau^\prime 
\frac{1}{1-K} (\tau_1,\tau_2, \tau, \tau^\prime)\mathcal{F}_0(\tau, \tau^\prime, \tau_3, \tau_4) .
\ee

A way to proceed is to diagonalize the rung operator $K$.
However if 1 is an eigenvalue of $K$, we will face a divergence and need to return to the full SD equations. 

Recalling the formula for the two point function in the approximate conformal (infrared) limit at zero temperature
$G_c ( \tau) = \frac{b }{\vert \tau \vert^{2 \Delta}} \sign \tau $
with $ b^q J^2 \pi = \left(\frac{1}{2}  - \Delta\right) \tan (\pi \Delta)$
we find that in this limit the kernel $K$ becomes (after symmetrizing with respect to $(\tau_1,\tau_2) \leftrightarrow (\tau_3, \tau_4))$
\bee K_c (\tau_1,\tau_2, \tau_3, \tau_4) = - \frac{1}{\alpha_0} 
\frac{\sign (\tau_{13}) \sign (\tau_{24}) }{\vert \tau_{13}\vert^{2 \Delta}
\vert \tau_{24}\vert^{2 \Delta} \vert \tau_{34}\vert^{2-4 \Delta}  }
\ee
\bee\alpha_0 = \frac{2 \pi q}{(q-1)(q-2) \tan (\pi / q)}
\ee

Conformal invariance allows us to simplify the problem by reexpressing $K$ as a function of the cross ratio
$\chi = \frac{\tau_{12} \tau_{34} }{\tau_{13}\tau_{24}}$
acting on single variable rung functions 
\bee \mathcal{F}_{n+1} (\chi) = \int \frac{d \tilde \chi }{\tilde \chi ^2} K_c (\chi, \tilde \chi)  \mathcal{F}_{n} (\tilde \chi) .
\ee
To further simplify the diagonalization, it is important to find out operators commuting with $K_c$. 
Recalling the $SL(2,\mathbb{R})$ invariance of the two-point function, the associated Casimir operator $C  = \chi^2 (1 - \chi) \partial^2_\chi  - \chi^2 \partial_\chi$ is such an operator, with a known 
complete set of eigenvectors $\Psi_h (\chi)$ with eigenvalues $h(h-1)$. They are therefore 
also the eigenvectors of $K_c (\chi, \tilde \chi) $. The strategy to compute $\mathcal{F} $
can then be roughly summarized as

\begin{itemize}
\medskip\item

Find properties of $\mathcal{F}_n (\tilde \chi)$ and	
the eigenvectors $\Psi_h (\chi)$ of the Casimir operator $C$
with these properties.

\medskip\item
Deduce conditions on $h$. One finds two families,
$h=2n$ with $n \in {\mathbb N}^\star$ and
$h = \frac{1}{2} + is$, $s \in {\mathbb R}$

\medskip\item
Compute the eigenvalues $k_c (h)$ of the kernel $K_c$
and the inner  products $<\Psi_h , \mathcal{F}_{0} >$ and 
$<\Psi_h , \Psi_h >$.

\medskip\item
Conclude that the 4 point  function is
\bee \mathcal{F} = \frac{1}{1-K} \mathcal{F}_0 = \sum_h \Psi_h (\chi )
\frac{1}{1 - k_c (h)} \frac{<\Psi_h , \mathcal{F}_{0} >}{<\Psi_h , \Psi_h >}
\ee

\medskip\item
But... \emph{one finds a single $h=2$ mode with $k_c (h) =1$}, which requires a special desingularization.
\end{itemize}
For $h = \frac{1}{2} + is$ or $h = 2n$ one can compute, for $\chi<1$,
\bee
\Psi_h = A \frac{\Gamma(h)^2}{\Gamma(2h)}\chi^h \tensor[_2]{F}{_1}(h,h,2h,\chi) + B \frac{\Gamma(1-h)^2}{\Gamma(2 - 2h)}\chi^h 
\tensor[_2]{F}{_1}(1-h,1-h,2-2h,\chi)
\ee
with 
$A = \frac{1}{\tan\frac{\pi h}{2}}\frac{\tan \pi h}{2} $ and $B = -\tan \frac{\pi h}{2} \frac{\tan \pi h}{2}$. For $ \chi >1$
\bea
\Psi_h &=& \frac{\Gamma(\frac{1-h}{2})\Gamma(\frac{h}{2})}{\sqrt \pi}\tensor[_2]{F}{_1}\left(\frac{h}{2},\frac{h}{2},\frac{2 - 2h}{2},\left(\frac{2 - 2\chi}{\chi}\right)^2\right),\\
\Psi_h &=& \frac{1}{2}\int_{-\infty}^\infty d y \frac{\abs{\chi}^h}{\abs{y}^{h}\abs{\chi - y}^{h}\abs{1 - y}^{1- h}}.
\eea
The conformal spectrum of $K$ follows as:
\bee
k_c(h)= -(q-1) \frac{\Gamma(\frac{3}{2} - \Delta) \Gamma(1 - \Delta)}{\Gamma(\frac{1}{2} + \Delta)\Gamma(\Delta)}\frac{\Gamma(\frac{h}{2}+ \Delta)}{ \Gamma(\frac{3 - h}{2} - \Delta)}\frac{\Gamma(\frac{1-h}{2} + \Delta)}{\Gamma(1 + \frac{h}{2} - \Delta)}.
\ee

\centerline{
\includegraphics[width=8cm]{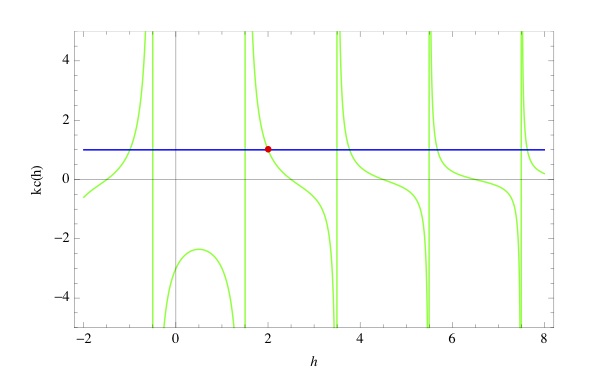}}

The conformal part of the 4-point function
\begin{align}
\frac{\mathcal{F}_{h\neq 2}(\chi)}{\alpha_0} 
=& \int_0^\infty \frac{\dd s}{2\pi}\frac{2h - 1}{\pi \tan(\pi h)} \frac{k_c(h)}{1 - k_c(h)} \Psi_h(\chi) + \sum_{n\geq 2} \left[\frac{2h-1}{\pi^2}\frac{k_c(h)}{1 - k_c(h)}\Psi_h(\chi)\right]_{h = 2n}\nonumber\\ 
=& -\sum_{m\geq 0} Res \left[\frac{h-1/2}{\pi\tan(\pi h/2)}\frac{k_c(h)}{1 - k_c(h)}\Psi_h(\chi)\right]_{h = h_m}\nonumber\\
=& \sum_{m\geq 0} c_m^2 \left(\frac{N}{\alpha_0}\right)\left[\chi^{h_m} \tensor[_2]{F}{_1}(h_m,h_m,2h_m, \chi)\right],\\ 
c_m^2 =& \frac{- \alpha_0}{N}\frac{h_m - 1/2}{\pi \tan(\pi h_m/2)}\frac{\Gamma(h_m)^2}{\Gamma(2 h_m)}\frac{1}{-k'(h_m)},
\end{align}
is written as a sum over conformal blocks, indicating the spectrum of operators of the model. 
However this regular contribution is subdominant in the IR regime. 
A careful treatment of the dominant, although conformally 
divergent mode $h=2$ will show that the MSS bound is indeed saturated. 

\subsection{The $h=2$ Divergent Mode}

To treat this divergent mode,
we need to compute the deviations to conformal
invariance at least to first order in $1/\beta J$. Corrections to the rung operator eigenvalues can be obtained as in time-independent perturbation theory in quantum mechanics. Anticipating the analytic continuation, it is convenient to work on the thermal circle: $\theta = 2\pi \tau/\beta$. Then, varying the conformal SD equations, one finds that reparametrizations of the two-point function are $K_c$-eigenfunctions with eigenvalue $k(h)= 1$ and of proper conformal weight $h=2$. For linearized reparametrizations $\theta \rightarrow \theta + \epsilon(\theta)$, written as $\epsilon_n = e^{-in\theta}$, the reparametrization modes $\Psi_{h=2}$ mode break into an infinite family
\begin{gather}
\Psi_{2,n} = \gamma_n \frac{e^{-iny}}{2\sin \frac{x}{2}}f_n(x), 
\quad f_n(x) = \frac{\sin \left(\frac{nx}{2}\right)}{\tan \frac{x}{2}} - n \cos \left(\frac{nx}{2}\right), \\
x = \theta_{12},  \quad y = \frac{\theta_1 + \theta_2}{2},  \quad \gamma_n^2 = \frac{3}{\pi^2 \abs{n}(n^2 - 1)}.
\end{gather}
At large $q$, an analytic expression for the $K$-eigenfunctions and their eigenvalues can be found for all couplings. Selecting among the first those that lead to the above functions in the IR, the associated eigenvalues are then used to get the first IR-corrections to the $K$-eigenvalues 
\begin{equation}
k(2,n) =  1 - \frac{3\abs{n}}{\beta \mathcal{J}} + \frac{7n^2}{(\beta \mathcal{J})^2} + \mathcal{O}\left((\beta \mathcal{J})^{-2}\right).
\label{eq:correction_k}
\end{equation}
Supported by numerical solutions, \cite{MS2016} extrapolated large $q$ results to estimate for finite $q$
\begin{equation}
k(2,n) = 1 - \frac{\alpha_K}{\beta \mathcal{J}}\abs{n} + \mathcal{O}\left((\beta \mathcal{J})^{-2}\right),  \qquad \alpha_K \equiv -q k^\prime(2) \alpha_G. 
\end{equation}
Plugging into ${\mathcal F}$ and taking down the extra two-point functions of the reparametrizations we get at leading order
\bea
\frac{\mathcal{F}_{h=2} (\theta_1, \theta_2, \theta_3, \theta_4)}{G(\theta_{12} )G(\theta_{34}) } = \frac{6\alpha_0}{\pi^2 \alpha_K}\beta \mathcal{J} \sum_{\abs{n}\geq 2}
\frac{e^{in(y^\prime - y)}}{n^2(n^2 - 1)}&\left[ \frac{\sin \frac{nx}{2}}{\tan \frac{x}{2}} - n \cos \frac{nx}{2}\right]
\left[ \frac{\sin \frac{nx^\prime}{2}}{\tan \frac{x^\prime}{2}} - n \cos \frac{nx^\prime}{2}\right].
\eea

For $\theta_1 > \theta_3 >  \theta_2> \theta_4$ :
\bee
\frac{\mathcal{F}_{h=2} (\theta_1, \theta_2, \theta_3, \theta_4)}{G(\theta_{12} )G(\theta_{34}) } =  \frac{6\alpha_0}{\pi^2 \alpha_K}\beta \mathcal{J} \left(\frac{\theta_{12}}{2\tan \frac{\theta_{12}}{2}} - 1 - \pi \frac{\sin \frac{\theta_1}{2}\sin \frac{\theta_2}{2}}{\abs{{\sin \frac{\theta_{12}}{2}}}}\right) ,
\ee
with $\theta_3 =\pi,\theta_4 =0$. For the ``regularized" OTOC \cite{Maldacena:2015waa}, $\theta_2 = \pi/2 - 2\pi i t/\beta = \theta_1 + \pi$
\bee
\frac{\mathcal{F}_{h=2} (\theta_1, \theta_2, \theta_3, \theta_4)}{G(\theta_{12} )G(\theta_{34}) } 
= \frac{6\alpha_0}{\pi^2 \alpha_K}\beta \mathcal{J} 
\left(1 - \frac{\pi}{2}
\cosh \frac{2\pi t}{\beta}
\right),
\ee
hence $\lambda_L = 2 \pi T $. In units where $\hbar =1$ this is the \emph{MSS bound}.

Other approaches leading to the chaos exponent are worth mentioning. For instance, correlation functions of the Fermions can be obtained from the (non-local) effective action of the two-point function and self-energy (see the subsection \ref{subsection : bilocal action}). A simpler way, if one is only interested in the chaotic regime, is to look for eigenfunctions of the ladder kernel (using analytically continued propagators) with eigenvalue $1$ and exponential in time (\cite{MS2016}, or \cite{Murugan:2017eto} for two-dimensional, Bosonic and supersymmetric variants). 

This large $N$ computation allows to see the initial growth in time of the correlation function. However one needs to sum the full perturbative series in order to see it reaching the equilibrium values. This is done in \cite{Maldacena:2016upp} from the bulk point of view (where the $1/N$ expansion becomes a $G$ expansion).

\section{Holographic Tensor Models}
\label{sec : holographic tm}

\subsection{The Three $1/N$ Expansions}

Up to now essentially three types of $1/N$ expansions have been discovered. The vector $1/N$ expansion is led by iterated tadpoles or \emph{cacti} graphs. The matrix $1/N$
expansion is led by \emph{planar} graphs \cite{tHooft:1973alw},
and is topological, governed by the genus of the underlying Riemann surface. The tensor $1/N$ expansion is led by \emph{melonic} graphs. The latter look much closer to cacti than to planar graphs although they are subtly different. However the main difference between vectors and tensors lies in the subleading terms: the $1/N$ tensor expansion has a much richer and more complicated  structure than its vector or matrix counterpart. As mentioned earlier, this structure at rank $d$ provides indeed a hierarchy for piecewise linear quasi-manifolds in dimension $d$.

Models with a quenched tensor saturated by vector variables were introduced much before SYK. The so-called $p$-spin glass model \cite{Gross1984} is a good example\footnote{Spin-glass tensor models with truly interacting tensors were introduced much later, in \cite{bgsspinglass}. The authors showed that, for a generalized p-spin model in which the disorder includes tensor interactions, in the large $N$ limit replica symmetry breaking still occurs, although at a renormalized temperature.}. Similarly, {\it melonic} graphs (a very natural family of parallel/series graphs) were certainly considered (under various names) by mathematicians and theoretical physicists much before the tensor $1/N$ expansion was found. 

The connection between disorder and random tensors is also not new. But the melonic limit for a \emph{single tensor saturated with vectors} is of a simpler nature than for \emph{truly interacting tensors}. Indeed when a single Gaussian random tensor saturated with vectors appears \emph{linearly} in the interaction, like an intermediate field, integrating on it provides a simple pairing of $q$-valent vector vertices into larger effective $2q$-valent vertices of the remaining \emph{vectors}\footnote{This pairing occurs both for quenched or annealed tensors, although the quenched models have a more complicated structure.}. The melonic dominance follows then simply from the \emph{cactus} dominance of that effective vector model; the melons are recovered when we reexpand the $2q$-valent effective vertices of the cacti into their initial pairs of $q$-valent components. For this reason the $1/N$ expansion of the SYK model is the one of a \emph{vector model in disguise}\footnote{From a more practical point of view \cite{Bonzom:2017pqs}, in a colored version of SYK, one can explicitly organise the perturbation theory in terms of an integer corresponding to a type of loops (their ``cyclomatic" number) whereas for the GW model (see the next subsection), one cannot escape that the expansion is governed by the Gurau degree (see subsection \ref{subsec:nlo}).}. 


It is now clear that many details in the SYK model are not essential (Bosons or Fermions\footnote{Actually the distinction is more subtle. Whereas the statistics doesn't alter the melonic dominance at large $N$, obtaining an interacting and conformal fixed point is not straightforward \cite{Murugan:2017eto}, one also needs to get around the spin glass phase at low temperature for Bosonic SYK \cite{Fu:2016yrv}. Then, combining Fermionic and Bosonic degrees of freedom led to interesting SYK-like behavior with supersymmetry (for instance \cite{Fu:2016vas} or \cite{Peng:2016mxj} using tensor models), or without (\cite{Peng:2017kro}).}, real or complex, particular rank, etc.). But an indispensable element for the saturation of the MSS bound and for the interesting gravitational aspects of the model is the melonic limit.  With hindsight we can even suggest that these gravitational aspects, which are different from those of a higher spin theory \cite{MalNati}, might be simply due to the coincidence of the $1/N$ SYK melonic limit with the one of a true tensor model. 

This suggestion can be explored more concretely by studying the difference between the subleading $1/N$ terms of the initial SYK model and of the more recent and truly tensorial Gurau-Witten (GW) \cite{Witten:2016iux,Gurau:2016lzk} and Carrozza-Tanasa-Klebanov-Tarnopolsky (CTKT) models \cite{Carrozza:2015adg,Klebanov:2016xxf}. 
They open the chapter of \emph{holographic tensors}, which we now describe. 


\subsection{The Gurau-Witten Model}

Late in 2016 E. Witten remarked the link between the SYK model and random tensors \cite{Witten:2016iux}. 
He proposed a modification to eliminate the quenched disorder, introducing the action
\bee I = \int dt  \biggl( \frac{i}{2} \sum_{i} \psi_i \frac{d}{dt}\psi_i  -i^{q/2}	j 	\psi_{0} \psi_1 \cdots \psi_{D}\biggr)
\ee
where $\psi$'s are $D+1$ real Fermionic tensors. The pattern of index contraction is 
the one of Gurau's initial colored tensor model, and
the $1/N$ complete expansion for this variant was established in \cite{Gurau:2016lzk}. 

The symmetry group is $O(N)^{D(D+1)/2}/\mathbb{Z}_2^{(D-2)(D+1)/2}$\footnote{To be compared to the $O(N)$ symmetry after disorder average of SYK.}. Indeed, in the interaction term, the contraction scheme associates pairs of the $\psi_i$ tensor indices, each transforming under $O(N)$. The quotienting invariant subgroup arises as the center of each $O(N)$ factor and imposing that it acts trivially on each Fermion $\psi_i$ of the interaction term.

Again let us stress the main difference between such an interacting tensor model and an SYK or a matrix model. In vectors and SYK models, the dimension of the symmetry group is larger than the number of fields as it scales as $N^2$ whether the fields have $N$ components. In matrix models, it is of the same order (both fields and symmetry scale as $N^2$). In tensor models, the symmetry is smaller than the fields: the symmetry still scales as $N^2$ whether the fields have $N^d$ components with $d \ge 3$. This is consistent with the idea that tensor models represent a higher level of abstract symmetry  breaking \cite{tensortrack} and a lower level of integrability, allowing for more chaotic dynamics.

Another important aspect is the possibility to gauge the symmetry and recover better understood features of holography (as the fundamental field of the boundary quantum theory has no clear bulk picture in known examples of the duality) \cite{ADSCFT}.

\subsection{The Carrozza-Tanasa-Klebanov-Tarnopolsky Model}

Following the trend towards simpler more economic 
tensor models that led from colored to uncolored tensors \cite{Bonzom:2012hw}, an  uncolored, i.e. single tensor model with similar properties, was soon developped
by Klebanov and Tarnopolsky \cite{Klebanov:2016xxf}, based on the three dimensional 
$O(N)$ tensor model interaction introduced by of Carrozza and Tanasa
\cite{Carrozza:2015adg}.

The Fermionic CTKT model is in some ways the simplest tensor model with SYK-like large-$N$ limit.
Its classical action is:
\be
{\bf S}_{\rm CTKT}[\psi] = \int \dd t \left( \frac{1}{2}  \psi_{abc} \partial_t \psi_{abc} + \frac{\lambda}{4N^{3/2}} \psi_{a_1 a_2 a_3} \psi_{a_1 b_2 b_3}  \psi_{b_1 a_2 b_3} \psi_{b_1 b_2 a_3} \right).
\ee

The contraction structure of the interaction vertex can be represented in the following way
\begin{figure}
\centerline{\includegraphics[width=14cm]{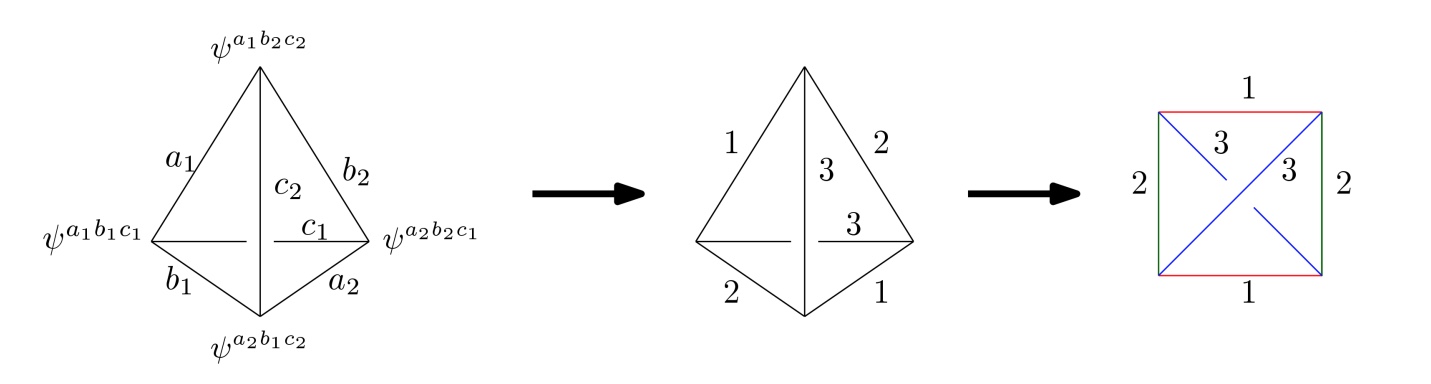}}
\caption{Tetrahedron Interaction}
\end{figure} 
whose dual graph is a tetrahedron. Higher rank versions of this model may be built using complete graph colorings, see \cite{Ferrari:2017jgw}.

\subsection{Main features}

In both models, analysis of the 2 and 4 point function led to the same SD equations than for SYK and implied consequently that they both saturated the chaos bound. 

However, the search for a comprehensive understanding of the theories, paralleled to tentatives to build their holographic duals, prompted studies in several directions. 

Firstly, as was done for SYK, where they managed to extract chaotic properties of the system from spectral information (for instance \cite{Cotler:2016fpe}), numerical computations at finite $N$ tackled the question of the energy spectrum and associated eigenstates\cite{Krishnan}. They looked for symmetries and degeneracies, and other properties (gaps, level repulsion, rigidity, etc.) pointing towards the chaotic ``dim-ramp-plateau''. This revealed that the statistics of the Hamiltonian of tensor models is governed by different ensembles than that of SYK. The biggest challenge is the extremely fast increase in the size of the Hamiltonian with $N$. 

With a similar point of view, \cite{Klebanov:2018nfp} started an analytical study at finite $N$ of CTKT and obtained (expected) bounds of the energy spectrum, the energy gap between singlet and non-singlet states (showing the the energy levels become dense, hinting at the conformal regime) and demonstrated the rapid growth with $N$ of the number of group invariant states. 

Secondly, and it was one the first things pointed out, \cite{Bulycheva:2017ilt} initiated the enumeration of multi-particle operators (more indices give much more freedom for contracting them) providing evidence for a Hagedorn transition\footnote{Briefly, an exponential behavior of the energy density makes the partition function singular.} in tensor models at a temperature of order $1/\log N$. Correlation functions of certain multi-particle operators contained also multi-ladder corrections, with an enhancing factor $\beta J$ for each ladder (cf. the 4-point function computation).

Thirdly, looking at the spectrum of conformal dimension of the bilinears in the Fermions, they noticed that it contained that of the complex or real version of SYK, depending on the symmetries imposed on the indices of the operator (antisymmetric or symmetric traceless respectively). Considerations of the spectrum of conformal dimensions of bilinears in the GW model above, as well as its diagrammatic expansion \cite{Bonzom:2017pqs}, suggest it to be the tensor equivalent of a ``4-flavour'' generalization of SYK \cite{Gross:2016kjj}.

Finally, important differences of GW with SYK were raised by \cite{Choudhury:2017tax} : it motivated the presence of $D\frac{N^2}{2}$ additional light modes, with a low energy action given by a $\sigma$-model, besides the conformal mode governed by the Schwarzian action as in SYK. An inquiry of thermodynamical aspects of the theory provided support for the aforementioned Hagedorn transition.

\subsection{Next-to-Leading-Order Contributions}
\label{subsec:nlo}

The leading $1/N$ contribution of the SYK and of holographic tensor models is the same, namely
it is given by the sum over melonic graphs. However the more promising tensor models have different 
subleading contributions. Whereas "vector-like" models such as SYK, although bilocal, have subleading
contributions essentially characterized by "excess", i.e. by the number of independent loops added to the tree
structure of the melons, true tensor models have  subleading contributions characterized by the Gurau degree, which is an average over the genera of \emph{jackets} which form a set of Riemann surfaces canonically embedded in the tensor graphs \cite{Ryan:2011qm}. 

In practice the fact that subleading contributions are different in SYK and tensor models has been first proved (with explicit examples treated in detail) in \cite{Bonzom:2017pqs}. Even though the same type of graphs appear (in the colored SYK model and GW), their scalings in $N$ are different and the tensor models, which contain more structure, lift the degeneracy of the SYK expansion. 
For example, in the case of the 4-point function, the Gurau-Witten
model distinguishes between broken and unbroken chains \cite{GS}, whether
the colored SYK model does not. This distinction becomes more and more important 
when digging further and further through subleading orders in the $1/N$ expansion of the two models, and means that the physics of the two models ought to be ultimately quite different.

\section{Further Issues}
\label{sec : further issues}
The holographic tensor models concealed in their simplicity a considerable richness. In order to clear a path towards their holographic dual, several directions have been proposed. Next, we will consider only a few of them. 

\subsection{Matrix-Tensor Models}

The tensor large-$N$ limit has also been used to derive a new large-$D$ expansion for multi-matrix models \cite{Ferrari:2017ryl}. The scaling in such models is borrowed from \cite{Carrozza:2015adg}, but 
the inspiration for this line of thought comes from works of Emparan et al. \cite{emparan}, studying general relativity in a large dimension $D$ expansion. In this ``mean-field'' approximation, the analysis of scattering modes on black hole backgrounds simplifies considerably due to the gravitational field of the black hole localizing close to the horizon. Furthermore, in contrast to most versions of tensor models, matrix models have a known interpretation in string theory.

In \cite{Ferrari:2017jgw}, the rescaling of \cite{Ferrari:2017ryl} was generalized.
Furthermore interactions based
on particular edge-colorings of the complete graph were studied at odd rank $D$ and it was discovered that the dominant melonic series has
standard combinatoric weights only for $D$ prime.

In \cite{Azeyanagi:2017drg} the  analysis of \cite{Ferrari:2017ryl}
is generalized to arbitrary multi-trace interaction terms and to arbitrary multiply-connected interaction bubbles. In \cite{Azeyanagi:2017mre} the  phase  diagram  of  this large $D$ limit is studied both in the Fermionic and Bosonic cases. 
In the Fermionic strongly coupled case there is a line of first order phase transitions which ends at a critical point which is studied numerically in detail. 

\subsection{Symmetric/Antisymmetric Tensors}

About one century ago, symmetry or antisymmetry on tensorial indices obviously played a great role in the differential and Riemannian geometric context which led to the formulation of general relativity by A. Einstein. Perhaps for this unconscious reason, the modern tensors of discrete simplicial geometry introduced to quantize gravity have been also often assumed totally symmetric in their indices, without further ado or justification \cite{earlytensors}. This is also the case in the princeps paper of Boulatov on group field theory \cite{Boulatov}. 

However the dominant graphs of fully symmetric tensor models contain tadpoles which are quite similar to the ultraviolet/infrared mixing terms in non-commutative field theory \cite{Minwalla:1999px}. They correspond to singular dual spaces, hence do not seem promising for quantum gravity applications. This difficulty seriously hampered for many years the development of tensor models. It was automatically solved by the introduction of \emph{colored} group field theories and tensor models, since the edge-colored condition automatically removes tadpoles \cite{Gurau:2010nd}.

Although the $1/N$ tensor expansion was then also established for unsymmetrized (i.e. colored) tensors, the question of how to generalize this expansion correctly to symmetric tensors is natural. In rank 2, this corresponds to the difference between Wishart (possibly rectangular) matrix models and the GOE or GUE ensembles for symmetric or Hermitian matrices.

It was conjectured in \cite{Klebanov:2017nlk} 
(supported by an exhaustive computer search of all rank 3 tensor graphs up to order 8)
\cite{Klebanov:2017nlk} that at rank three the correct ensemble of tensors for which the melonic sector leads to an interesting $1/N$ expansion should be the ensemble of \emph{traceless symmetric} tensors (i.e. for which $\sum_a T_{aab} =0 \; \forall b$), or 
the ensemble of \emph{totally antisymmetric} tensors. Indeed any of these two conditions prevents the formation of the undesired tadpoles.

This conjecture has now been proved for rank 3 tensors in two steps. In \cite{Gurau:2017qya} it was proved for a bipartite model made of two symmetric tensors
(the bipartite condition preventing tadpoles). In \cite{Benedetti:2017qxl} the conjecture was fully proved at rank three both for the traceless symmetric and the antisymmetric case, using a detailed but tedious case by case analysis. More recently the last irreducible
$O(N)$ case at rank three, namely the mixed permutation symmetry case, was also solved 
\cite{Carrozza:2018ewt}.

The next step is now clearly to 
extend these results at higher rank $D>3$. This seems to require a general method to avoid a case by case analysis, and we suspect that this fascinating program will be full of unexpected twists and discoveries.

\subsection{Probes and coupling to lower ranks}

It is tempting to couple the tensors to different types of fields and see under what conditions we recover chaos. This was the motive of \cite{Peng:2017kro}. From an interaction between Fermionic tensor and vector fields $\psi_{abi}\psi_{abj}\chi_i\chi_j$, where no chaos was found, a second term $\psi_{abi}\psi_{acj}\phi_b\phi_c\chi_i\chi_j$ involving a Bosonic vector was added. This time, depending on the sign of the additional interaction, one observes the saturation of the chaotic bound\footnote{And it doesn't depend on the statistics of the tensor field.}. 

From another perspective, it was noticed \cite{Halmagyi:2017leq} that one can remove lines of the exterior fields in the stranded representation of melonic graphs without affecting their power counting. This could be pictured as an additional coupling of tensors $\kappa$ of lower rank to the ones of rank $D$ $\psi$: $\psi_{abc}\psi_{ade}\kappa_{bd}\kappa_{ce}$. As the former don't affect the leading order correlation functions of the latter, they are interpreted as probes for the background generated by the larger rank tensors, in the same fashion as in \cite{IP-IOPmodel}, vectors were probing the black hole background generated by matrices\footnote{However those don't feature chaos \cite{Michel:2016kwn}.}. Again computing OTO four-point function involving different sets of fields, it was verified that $\expval{\psi^4}, \expval{\kappa^4}, \expval{\psi^2 \kappa^2}$ all present exponential growth\footnote{Although the mixed correlation functions seem to grow faster than the background fields.}. It was also indicated that including a vector as well with $\psi_{abc}\kappa_{ad}\kappa_{bd}\eta_c$ and a colored version of the model would demonstrate the same chaotic features.

\subsection{Bilocal Action}
\label{subsection : bilocal action}

Taking an annealed average of the disorder in the SYK model (assuming replica diagonal solution) led to a bilocal action in terms of the 2-point function of the Fermions and their self-energy. This allowed also to extract a global factor of $N$ in front of the bilocal action and study its $1/N$ expansion as a saddle-point analysis \cite{MS2016,Kitaev:2017awl}.

This procedure has also been developed within collective field theory \cite{Jevicki}. Furthermore, it connects to the effective action of reparametrization modes of the boundary of a regulated $AdS_2$ \cite{Maldacena:2016upp}. A dilaton is used to interpolate between the $AdS_2$ throat of the near-horizon region of a near-extremal black hole and the remaining spacetime, breaking explicitly the $SO(2,1)$ symmetry, similarly to the derivative term breaks the diffeomorphism invariance of SYK\footnote{The lectures \cite{Gabor} give an excellent summary of the connection.}. 

Using the formalism of the 2PI-effective action, a similar point of view was reached in \cite{Benedetti:2018goh}: the authors derived a bilocal action for the above tensor models. In practice the method consists in taking a Legendre transform of a modified generating function, including beside the current coupling to the field, a source coupling to a bilinear in the field\footnote{We used repeated indices to represent summations and/or integrations.}
\bea
W[J,K] &=&\ln \int \mathcal{D}\varphi \exp \left(-S[\varphi] +J_a \phi_a +  \frac{1}{2}\phi_a K_{ab} \phi_b\right),  \\
\Gamma[\phi,G]&=& - W[J,K] + \fdv{W}{J_a}J_a + \fdv{W}{K_{ab}} K_{ab} = - W[J,K] + J_a \phi_a + \frac{1}{2} \phi_a K_{ab} \phi_b+\frac{1}{2}G_{ab}K_{ab},
\eea
The following relations
\bea
\Phi_a[J,K] = \fdv{W}{J_a}[J,K] \quad G_{ab}[J,K] = \frac{\delta^2 W}{\delta J_a \delta J_b}[J,K],
\eea
and others analogous with $\Gamma$ exchanging sources with fields, assumed invertible for $(J,K)$ and $(\Phi,G)$ respective inverse, allow to go from one set of variables to the other.
Denoting the free propagator $G_0^{-1} = \fdv[2]{S}{\phi}$, we obtain in a loop expansion (expanding the field $\varphi$ around its expectation value $\varphi = \phi + f$)
\bea
\Gamma[\phi,G]= S[\phi] + \frac{1}{2} \Tr[\ln G^{-1}] + \frac{1}{2} \Tr[G_0^{-1} G] +  \Gamma_2[\phi,G].
\eea The first terms in the last equality come from quadratic terms in $f$ and $\Gamma_2$, containing higher orders, is the generating function of 2PI graphs\footnote{Graphs that stay connected cutting 2 lines.}. Then, the equations of motion for $\phi$ and $G$ obtained from the effective action can be studied in a $1/N$ expansion keeping the relevant graphs.  

Applying this formalism to SYK, \cite{Benedetti:2018goh} reproduced at leading order in $1/N$ the SD equations of SYK and at next-to-leading order what could be seen as arising from integrating out a quadratic field with the same covariance as in \cite{Jevicki, Kitaev:2017awl} (however, it is inaccurate to extend this analysis at following orders : disorder averaging at the beginning in order to get equations of motions or disorder averaging at the end is different, in the same way that at subleading orders, the different replicas interact and a replica-diagonal ansatz in not enough). 

Similarly, the authors of \cite{Benedetti:2018goh} studied the CTKT and GW models. For the first, they rewrote the effective action as a gauge invariant term ($O(N)^3$ symmetry) and corrections taking the form of a non-linear $\sigma$-model as conjectured by \cite{Choudhury:2017tax}. The second one features a $\Tr\log$-type effective action (up to NNNLO), with a quite explicit distinction of the fluctuating fields at each considered order\footnote{The analysis was made easier by \cite{Bonzom:2017pqs} where the graph structure of the following orders in $1/N$ was detailed.}.

\subsection{Higher Dimensions}

To raise the NAdS$_2$/NCFT$_1$ SYK holographic properties to higher dimension
does not seem easy because as explained in 
\cite{Murugan:2017eto}, the most natural two-dimensional models (e.g. for Fermions)
are just renormalizable. The non-trivial infrared anomalous dimension of the
SYK model was due to the non-renormalizable infrared power counting and is no longer there when raising the dimension; also the MSS bound is no longer saturated. Nevertheless several papers have studied generalization of SYK-type tensor models to higher dimensions, and we shall 
briefly mention only three attempts.

One such paper \cite{Prakash:2017hwq} studied a $d$-dimensional version of the CTKT model. Given the similar melonic dominance in the correlation functions, they  studied the spectrum of bilinear operators in the Fermions and found that in $d<2$ one had a seemingly well-defined IR fixed point, between $2<d<6$ the spectrum contained a complex eigenvalue signaling a non-unitary theory and numerical results for $6<d<6.14$ indicated a real spectrum but still the possibility of having a non-unitary theory. 

In a similar perspective, for (quartic) Bosonic tensor models in $d = 4 - \epsilon$ dimensions, \cite{Giombi:2017dtl} showed that the conformal dimension of bilinears was complex. For higher rank interactions, they showed the existence of a dimension below which the quadratic operators aquired a complex dimension (and for a narrow range below $d=6$, their conformal dimension was real). 

In another paper  \cite{Benedetti:2017fmp}, the authors considered the two-dimensional (real and complex) tensorial version of the Gross-Neveu model (motivated by SD equations where free propagator and self-energy compete). However they could not find a stable and conformal IR fixed point. As a mass was dynamically generated in the complex case, breaking the continuous $U(N)^3$ symmetry, they also explain the mecanism of symmetry restoration, carefully obtaining the low energy effective action of the would-be Goldstone modes and motivating the form of their two-point function as showing at large distances exponential decay at finite $N$ but quasi-long-range order (i.e. decaying like $\abs{x}^{-c/N^\alpha}$ for $c$ and $\alpha$ positive constants) doing first large $N$. In the real case, they could show the presence of a weakly interacting IR fixed point in $2-\epsilon$ dimensions, the tetrahedral interaction being marginally irrelevant in exactly two dimensions.

\subsection{Tensor Field Theories}


TFTs \cite{TFT,TGFT} distinguish themselves from tensor models and from ordinary group field theories by the presence of a non-trivial propagator, which may or may not include a gauge projection of the Boulatov type \cite{Boulatov}. Typical TFT's interactions remain invariant
but the propagator (i.e. the Gaussian measure covariance)  is purposefully chosen to slightly break the tensor symmetry\footnote{
This is quite natural if we consider the tensorial symmetry as a kind of abstract generalization 
of  \emph{locality} in field theory \cite{tensortrack}. Propagators, as their name indicates, should \emph{break} locality.}. This allows to exchange the $1/N$ limit for the physically more familiar picture of power counting and renormalization group analysis.

Indeed the main consequence of this slight breaking of the tensor symmetry is to allow for a separation 
of the tensor indices into (abstract, background-independent) ultraviolet and infrared degrees of freedom. 
Like in ordinary field theory most of the indices should have small covariances. They can be identified with the (abstract) ultraviolet
degrees of freedom of the theory. Therefore they should be integrated to compute the effective theory for the few indices which form the infrared, effective degrees of freedom (not the other way around!). Remark that this picture seems compatible with the general AdS/CFT philosophy in which the renormalization group time, which flows between different conformal fixed points,
provides the extra bulk dimension of AdS \cite{Ramallo:2013bua}.

At rank 2, TFTs reduce to non-commutative quantum field theory (NCQFT), which is an effective regime of string theory \cite{Douglas:2001ba}. Mathematically it 
corresponds to matrix models of the Kontsevich-type \cite{Kontsevich:1992ti}. One of the most studied such theory, the Grosse-Wulkenhaar model, can be renormalized \cite{Grosse:2004yu}. The leading planar sector displays beautiful features such as 
asymptotic safety \cite{Disertori:2006nq}, integrability \cite{Grosse:2012uv} and unexpected restoration of Poincar\'e symmetry and of Osterwalder-Schrader positivity \cite{Grosse:2016qmk}. 

At rank higher than two an important unexpected property of TFTs is their generic asymptotic freedom, at least for quartic melonic interactions \cite{BenGeloun:2012pu,AF,Rivasseau:2015ova}. Their renormalization group flow in the \emph{tensor theory space} \cite{Rivasseau:2014ima} can be investigated via functional equations and truncations to search numerically for non-trivial fixed points \cite{Benedetti:2014qsa}.

Until recently TFT's were equipped with inverse Laplacian propagators (as advocated in e.g. \cite{Geloun:2011cy}), possibly raised to some non-trivial power \cite{BenGeloun:2017xbd}. Also the TFT formalism had not been applied to SYK and holographic tensors, in which the tensor indices are abstract and the $N \to \infty$ limit is performed at the start. 
Such an \emph{ab initio} $1/N$ limit cannot couple to the conformal limit, and this seems somewhat unphysical to us.  

Therefore it was proposed recently to investigate TFT's  equipped with interactions of the CTKT type and condensed matter propagators \cite{BenGeloun:2017jbi}. Indeed condensed matter models have the interesting property that their infrared power counting, governed by the Fermi surface singularity, is always just renormalizable, independently of the number of space dimensions. Their (dynamical) mean field theory occurs therefore at 
infinite dimension. This leads to some hope
that they could be helpful in order to build holographic tensor models in arbitrary dimension and to understand their mean field theory limit.

\medskip\noindent
{\bf Acknowledgments}

We thank the organizers of the Corfu summer school for a very enjoyable school. V.R. especially thanks G. Zoupanos for his remarkable tenacity in developing the Corfu summer schools. We are grateful to D. Benedetti for many insightful discussions, and to D. Benedetti and A. Tanasa for comments on the draft. We are also indebted to R. Pascalie for sharing with us his internship report on the SYK model and to J. Ben Geloun, R. Gurau and R. Toriumi for discussions on various aspects of these lectures.

\end{document}